\documentstyle[12pt]{article}

\newcommand{\beq}{\begin{equation}}
\newcommand{\eeq}{\end{equation}}
\newcommand{\bea}{\begin{eqnarray}}
\newcommand{\eea}{\end{eqnarray}}
\begin{document}

\date{}
\title{From Brans-Dicke theory to Newtonian gravity}
\author{Sergey Kozyrev\thanks{
e-mail : {\ Sergey@tnpko.ru}} \\
\\
Scientific center gravity wave studies "Dulkyn".}
\maketitle

\begin{abstract}
We present the new interpretation of scalar field for the
Brans-Dicke theory. This interpretation is obtained by considering
a fixed spacetime structure of manifold.
\end{abstract}


\section{Comparisons with Newtonian gravity}

The scalar-tensor theory first time was invented by P. Jordan
\cite{Jordan} in the 1950's, and then taken over by C. Brans and
R.H. Dicke \cite{Dicke} some years later. In this paper we
restrict our discussion to the Brans-Dicke theory \cite{Jordan,
Dicke}  that, among of all the alternative theories of classical
Einstein's gravity, is the most studied and hence the best known
theory.

 Scalar-tensor theories of gravity describe the universe
as grounded on differentiable arbitrary manifold $M^4$ enveloped
by a principal bundle formed with isometric representations of a
finite continuous Poincar\'{e} group. Einstein's principle of
general relativity asserts the invariance under general coordinate
transformations of the actions integral grounded on a $M^4$
manifold parameterized by variables $x^\mu, \mu = 0,1,2,3$.

As it is well known field equation and conservation low of the
relativity theory can be obtained from principle of least action.
The same principle is the basis of the Brans - Dicke theory
\footnote{Units $8\pi G=c=1$ are used throughout the paper. Greek
indices range over the coordinates of the 4-manifold and Roman
indices over the coordinates of the 3-surfaces.} \beq S = {S}_G
+{S}_m \eeq \beq\label{action} S_G =\frac12\int dx^4
\sqrt{-g}\left\{\Phi R -
\omega\frac{\Phi^{,\mu}\Phi_{,\mu}}{\Phi}\right\} +{S}_m, \eeq
where $R$ is the scalar curvature, $\phi$ is a scalar field,
$\omega$ is a dimensionless coupling parameter, and ${S}_m$ is an
action of ordinary matter (not including the scalar field).

An unsatisfactory feature of relativistic theories is that the
components of Lagrangian do not have any direct physical
interpretation. Note that in abstract manifold, the Ricci scalar
and tensor $g_{\mu\nu}$ lose their geometrical meaning that they
had in a spacetime and now can be viewed only as a "source" for
the metric.

The scalar - tensor theories appears as a theory in which the
gravity is described simultaneously by two fields the metric
tensor and the scalar fields, the latter being an essential part
of the geometrical property of spacetime  manifesting its presence
in all geometrical phenomena, such as curvature, geodesic motion
and so on.

On another hand, the Newtonian description of gravitating systems
was developed in the 17th century using a scalar potential field
and is nowadays a part of most classical mechanics textbooks. It
seems reasonable to interpret Newton's "absolute space" as an
absolute Euclidean embedding-space that acts as a container for
non-Euclidean geometry. But there may be as well  other reasons to
contemplate Minkowski space from considerations of scalar gravity.

In a gravity theories, a model of spacetime is usually a pair $(M,
T)$ where $M$ is $N$-dimensional manifold with suitable
topological and differentiable properties and $T$ represent a
collection of matter fields on  $M$. In approach with fixed
spacetime structure, like Newtonian mechanics and special
relativity, this model suggest an interpretation of manifold $M$
as independently existing "container" for the histories of fields
and particles. Obviously, the action (\ref{action}) can be endowed
with a structure of a manifold. To obtain the physical
interpretation of the scalar field, one may write the action
(\ref{action}) for Minkowski metric in the following manner:
$g_{\mu\nu}=\eta_{\mu\nu},$  $ R=0 $. We need field equations for
$\phi,$ so the action (\ref{action}) for this field must be
supplied,

\begin{equation}   \label{eacgrn}
  S_G
  =\int dx^4 \,\eta^{\mu\nu}\partial_\mu\phi \partial_\nu\phi,
\end{equation}
where $\phi = e^{\Phi}$. Without losing generality, suppose that
$\omega=1$.

 The overall action for the aggregate of N point
particles is

\begin{equation}                                                  \label{eacgpo}
  S_{ m}=-\sum_{i=1}^n\int_{-\infty}^\infty\!\!\!dt\,
  \frac12m_i\dot q_i^\mu\dot q_i^\nu\eta_{\mu\nu}
  -\sum_{i=1}^n\int_{R^4}\!\!\!dx\,m_i\delta(\textbf{x}-\textbf{q}_i)\phi.
\end{equation}
where $q_I(t)=\{ q_I^\mu(t)\},I=1,...,N$ are the trajectory of
point particles with mass $m_I$.

 As in Newtonian mechanics, we can consider arbirtary
spacelike section $t = const$ given by Euclidean metric. The
action (\ref{eacgrn}) provides the following field equations
\cite{Katanaev}:

\begin{equation}\label{eqmogr}
  \frac{\delta S_G}{\delta\phi} =\triangle\phi
 =0,
\end{equation}
where $\triangle:=\partial_1^2+\partial_2^2+\partial_3^2$ is a
three dimensional Laplasian, the gradient of scalar field
$\partial \phi$ is taken at the point $q_I = \{t, q_I ^\mu\} $,
where at the time the particle is located.

The field equations read

\begin{equation}                                                     \label{eqmogr}
  \frac{\delta S}{\delta \phi} = \frac1{4\pi G}\triangle\phi
  -\sum_{i} m_\delta(\textbf{x}-\textbf{q}_i)=0,
\end{equation}

\begin{equation}                                                     \label{eqmpon}
  \frac{\delta S}{\delta q_i} = m_i\left(\ddot q_{i\mu}-\partial_\mu\phi\right)=0,
\end{equation}

Then, with this definition, the equation of gravity field have the
form

\begin{equation}                                                     \label{epoise}
  \triangle\phi=\sum_I m_I\delta(\textbf{x}-\textbf{q}_I).
\end{equation}

Thus, we see that the scalar field variables of Brans-Dicke theory
play the role of a standard Newtonian potential.

 The field equation
(\ref{epoise}) yields the following solution \cite{Vladim}
\begin{equation}                                                  \label{esolpo}
  \phi(t,\textbf{x})=-\sum_I\frac{m_I}{|\textbf{x}-\textbf{q}_I(t)|},
\end{equation}
where
\begin{equation}
  |\textbf{x}-\textbf{q}_I|:=\sqrt{-\eta_{\mu\nu}(x^\mu-q_I^\mu)(x^\nu-q_I^\nu)}
\end{equation}

Now we consider the model with two particle with masses $m_I$ and
$m_J$. The force $F = {F^\mu}$, acting on a particle $m_I$  by
particle $m_J$, is
\begin{equation}
 \label{enewlo}
  F^\mu=m_I m_J\frac{q_I^\mu-q_J^\mu}{|\textbf{q}_I-\textbf{q}_J|^3}.
\end{equation}
\emph{It is Newton's law of gravitation.}

\section{Discussion}

 In order to make a physical prediction from theory a manifold must be endowed with a structure (metric, connection, curvature ...) In contrast with usual formalism of general relativity one can perform a model of spacetime with fixed background geometry, like Newtonian mechanics and special theory of relativity. Then the set up of action implies the imposing the Ricci scalar not as a scalar curvature of spacetime but as a matter source.
 The standard derivation of the Newtonian gravity as a weak field limit of relativistic theories do not expose the specific feature of Brans-Dicke theory which include Newton's law of gravitation as an exact solution.

\end{document}